\documentclass[useAMS,usenatbib,usegraphicx]{mn2e}

\newcommand{\simgt}{\;\hbox{\rlap{\raise 0.425ex\hbox{$>$}}\lower 0.65ex\hbox{$\sim$}}\;}
\newcommand{\simlt}{\;\hbox{\rlap{\raise 0.425ex\hbox{$<$}}\lower 0.65ex\hbox{$\sim$}}\;}

\title[Weak lensing of SNIa]
      {Weak lensing of the high redshift SNIa sample}
\author[L.L.R. Williams and J. Song]
       {Liliya L.R. Williams$^{1}$\thanks{E-mail:
        llrw@astro.umn.edu (LLRW); jsong4@astro.uiuc.edu (JS)} 
        and Jeeseon Song$^{1}$\footnotemark[1]\thanks{Present address:
Department of Astronomy 133 Astronomy Building, 1002 W. Green St. Urbana, IL 61801, USA}\\
$^{1}$Astronomy Department, University of Minnesota, Minneapolis, MN 55455, USA}

\begin{document}  



\maketitle

\label{firstpage}

\begin{abstract}
The current sample of high-redshift Supernova Type Ia, which combines results
from two teams, High-$z$ Supernova Search Team and Supernova Cosmology Project,
is analyzed for the effects of weak lensing. 
After correcting SNe magnitudes for cosmological distances, assuming recently 
published, homogeneous distance and error estimates, we find that brighter SNe 
are preferentially found behind regions (5-15 arcmin radius) which are 
overdense in foreground, $z\sim 0.1$ galaxies. 
This is consistent
with the interpretation that SNe fluxes are magnified by foreground galaxy
excess, and demagnified by foreground galaxy deficit, compared to a smooth
Universe case. The difference between most magnified and most demagnified
SNe is about 0.3-0.4 mag. The effect is significant at $>99\%$ level.
Simple modeling reveals that the slope of the relation between SN 
magnitude and foreground galaxy density depends on the amount 
and distribution of matter along the line of sight to the sources, 
but does not depend on the specifics of the galaxy biasing scheme. 
\end{abstract}


\section{Introduction}

The effects of weak gravitational lensing by the large-scale structure 
have been detected in several samples of high redshift QSOs, intermediate redshift
galaxies, and BATSE GRBs. In the case of point sources, QSOs and GRBs,
weak lensing manifests itself as angular (anti-)correlations between
these sources and foreground inhomogeneously distributed mass 
\citep{wf03,bsm01,wi98,bmg97},
while in the case of galaxies weak lensing is detected through its 
coherent shear effect (see \cite{ref03} for a recent review).
In principle, there is another, more direct way of detecting weak lensing,
which uses fluxes of standard candles. If the
observed magnitudes of standard candles are corrected for cosmological distances
then the effect of lensing can be seen: brighter sources will 
lie behind regions of mass density excess, while fainter ones will have
mass deficits in their foregrounds.

The best example of cosmological standard candle,  
Supernovae Type Ia (SNIa) have been extensively observed with the purpose of
determining the global geometry of the Universe 
\citep{tonry03,perlm99,gar98,riess98}.  Nuisance effects like 
evolution, variations in individual SN, and gray dust extinction have been
studied theoretically and observationally, and have either been corrected
for or shown to be small. Weak lensing, another nuisance effect has been 
addressed theoretically by several authors \citep{amg03,met99,holz98,wcxo97}
and found to be unimportant given the current uncertainties.
For example, ~\citet{wcxo97} used ray tracing through cosmological 
simulations and found that the lensing induced dispersions on truly standard 
candles are $0.04$ and $0.02$ mag at redshift $z=1$ and $z=0.5$, respectively, 
in a COBE-normalized cold dark matter universe with $\Omega_m=0.4$, 
$\Omega_\Lambda=0.6$, $H_0=65$km/s/Mpc and $\sigma_8=0.79$. 
These are small variations compared to the current errors
which are $\simgt 0.2$ mag.  Even though weak lensing effects are estimated to be 
small for $z_s <1$, they are predicted to be non-negligible for higher 
redshift sources, so it is not
surprising that the highest redshift SNIa, SN1997ff at $z_s=1.755$ has been 
examined by several authors \citep{ben02,mgg01,li01} 
for the effects of weak lensing due to galaxies along the line of sight.

Present day high-$z$ SNIa samples are dominated by lower redshift SNe, and
so have not been examined for the effects of lensing.
The main goal of this work is to determine if the observed fluxes of the 
cosmologically distant SNIa have suffered significantly from lensing induced 
(de-) amplifications.

\vfill

\section{Data}\label{data}

The largest homogeneous compilation of SNIa has been recently published
by ~\citet{tonry03}: Table 15 of that paper contains
74 SNe at $z_s\ge 0.35$ . The authors use four different light curve
fitting methods 
(MLCS, $\Delta m_{15}(B)$, modified dm15, and Bayesian Adapted Template Match)
to estimate distances to SNe. The final quoted distance is the median of the
estimates of the four individual methods, and the uncertainty is the median of
the error of the contributing methods.
The analysis presented in \citet{tonry03} yields values of the global
cosmological parameters; if a flat model is assumed, then $\Omega_m=0.28$ and 
$\Omega_\Lambda=0.72$. We use these values in all the analysis of the present
paper.

As tracers of foreground mass density we use APM galaxies \citep{irwin94}. 
APM provides near full coverage of the sky in the Northern and
Southern Hemispheres, at $|b|\simgt20^\circ$. In our analysis 
we use only the central $r=2.7^\circ$ of APM plates. 
Since the plate centres are separated by $\sim 5^\circ$, there exist small 
portions of the sky that are not covered by any plate. As a result of these cuts, 
only 55 of the 74 SNe lie on the usable parts of APM plates.

The median redshift of the 55 SNe is $0.47$.\footnote{SN1997ff at $z_s=1.755$
is not in our sample: it fell in the cracks between the APM plates.}
Since most of the SNe have rather low redshifts, care must be taken to 
ensure that galaxies are foreground to the SNe. Furthermore, because SNe
span a large range of nearby redshifts, from $z_s=0.35$ to $1.2$, the optimal
lens redshift $z_l$ will depend on $z_s$ much more compared to a typical case
where sources (say, QSOs) are at $z_s\sim 1-3$ and so the redshift  of optimal
lenses is roughly independent of $z_s$. In our analysis we adjust $z_l$ for each 
SN source by selecting the appropriate limiting apparent magnitude, 
mag$_{\;gal\;lim}$ for APM galaxies on red plates.
\citet{maddox96} gives an empirical expression for the median redshift
$z_{med}$ of a galaxy sample with a given faint magnitude flux cutoff. 
This median redshift can be equated with the optimal lens redshift 
$z_l$, and hence the magnitude limit of the foreground galaxies 
can be determined for every SN separately. However, there is a small catch.
For $z_s=0.4$ optimal $z_l=0.174$. The galaxy redshift distribution 
whose median redshift $z_{med}=0.174$ 
has a considerable tail extending beyond $z=0.4$. To avoid the problem
of source/lens redshift overlap we use $z_{med}=z_l/2$, where factor of 2 was 
chosen arbitrarily. We explore the dependence of the results on this factor 
in Section~\ref{robust}.

\section{Analysis}\label{anal}

Around every SN we draw a circle of radius $\theta=10^\prime$, and count 
the number of galaxies, $n_{gal,D}$, in the appropriate magnitude range.
This number is compared to the average number density in
control circles, $\langle n_{gal,R}\rangle$. Fractional galaxy excess
is $\delta n_{gal}=n_{gal,D}/\langle n_{gal,R}\rangle-1$. Control
circles are confined to the same APM plate as the SN, and to the same
distance from the plate centre as the SN (to reduce the effects of vignetting); 
however, scattering the control circles randomly on the plate does 
not change the results significantly. For each SN
we also calculate $N_{>}/N$, where $N_{>}$ is the number of control circles, 
out of total $N$, that have less galaxies in them than the circle around 
the SN. In other words, $N_{>}/N$ is the rank of the SN circle among its 
control `peers'. If SNe are randomly distributed with respect to the 
foreground galaxies, then average $\langle N_{>}/N\rangle=0.5$. 
If SNe have an excess (deficit) of galaxies in front of them then their 
$\langle N_{>}/N\rangle$ will be greater (less) than 0.5. Analogous to the 
medians being more stable than averages, $N_{>}/N$ rank statistic is more 
stable than $\delta n_{gal}$.

\begin{figure}
\includegraphics[width=84mm]{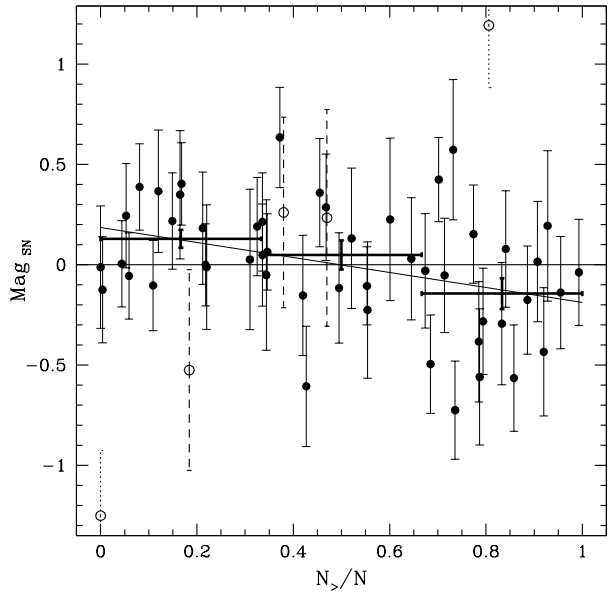}
\caption{Absolute magnitudes of SNIa (modulo a constant offset) versus the 
foreground galaxy density. Magnitudes were obtained assuming $\Omega_m=0.28$, and 
$\Omega_\Lambda=0.72$, and extinction, K-corrected distances from \citet{tonry03}.
$N_>/N$ is a measure of the number density of foreground galaxies in circles of 
radius $\theta=10^\prime$ around SNe (see Section~\ref{anal} for details).
There are a total of 55 sources (with magnitude errors), but only 50 
(filled points) are used in the analysis. The 50 points are grouped into three
bins, whose horizontal size is shown as thick horizontal ``error-bars''. The 
corresponding vertical error-bars show the deviation of the mean of the points 
in each bin (the rms is 4 times larger), and the intersection of the thick 
lines are the averages of the 
Mag$_{SN}$ of SNe in each bin. The thin slanting line is the best-fit to 50 SNe, 
and has a slope $\beta=-0.373$. The significance of the correlation is $>99\%$.}
\label{figone}
\end{figure}

Figure~\ref{figone} shows absolute magnitudes, $M_{SN}$, and $N_{>}/N$ ranks 
of 55 SNe found on APM plates. The effect of flux dimming due to cosmological 
distances has been taken out, i.e. all the SNe have been `brought' to the same
redshift; the constant magnitude offset on the vertical axis is irrelevant for
this work. There are two SNe whose magnitudes make them $>3\sigma$ outliers, 
SN1997O, and SN1997bd represented by empty circles with dotted line error-bars. 
We exclude these from our analysis.\footnote{Of our 55 SNe, SN1997bd has the 
highest host-galaxy extinction, $\langle A_v\rangle$ as quoted in \cite{tonry03}. 
SN1997O and SN1994H were excluded from the primary fit (Fit C) by the 
analysis of \citet{perlm99}. We do not exclude SN1994H from our analysis,
but if we did it would improve the trend, as its coordinates on Fig.~\ref{figone}
are (0.427; -0.606).}

\begin{figure}
\includegraphics[width=84mm]{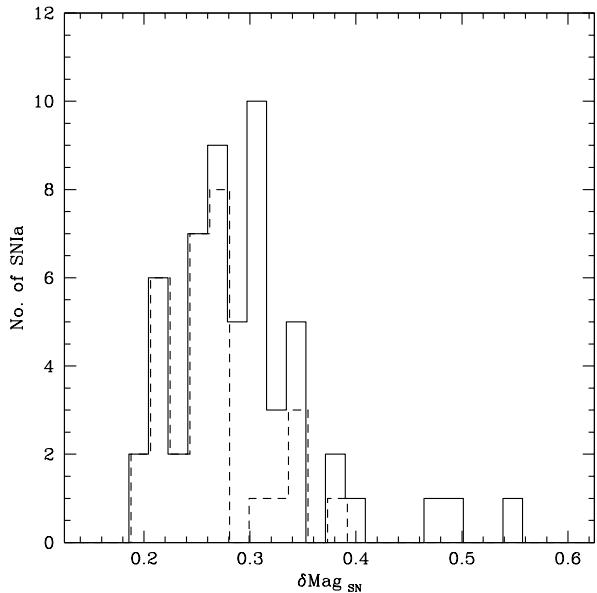}
\caption{Histogram of SNIa magnitude errors, $\delta M_{SN}=5\;\delta \log (d)$.
Distance errors, $\delta \log (d)$ were derived by 
\citet{tonry03}, and listed in column 9 of their Table 15.
The solid histogram is for 55 SN; the dashed line represents 31 SN from the
High-$z$ Supernova Search Team. The 3 outlier SNe with $\delta M_{SN}>0.45$ 
are not used in the present analysis.}
\label{figthree}
\end{figure}

The distance (or, magnitude) errors, as estimated by \citet{tonry03} for 
the 55 SNe are shown in Fig.~\ref{figthree}, as the solid line histogram. 
The dashed line represents the subsample of 31 SNe from High-$z$ Supernova 
Search Team \citep{tonry03}, whose errors appear to be generally smaller 
than those of the Supernova Cosmology Project \citep{perlm99}. We use SNe 
from both the groups, but exclude three outliers in Fig.~\ref{figthree},
whose $\delta M_{SN}>0.45 $. In Fig.~\ref{figone} these are represented by 
empty circles with dashed line error-bars. Thus, we exclude a total of 5 SNe 
from our analysis, leaving us with 50, shown as the solid points in 
Fig.~\ref{figone}. 

These 50 SNe exhibit a relation between their $M_{SN}$ and $N_{>}/N$, 
in the sense that brighter SNe have an excess of galaxies in their foregrounds. 
For illustration purpose only, we bin the 50 SNe into three bins; Fig.~\ref{figone}
shows the extent of the bins and the deviation of the mean of the SNe magnitudes
in each bin as thick lines.  

\begin{figure}
\includegraphics[width=84mm]{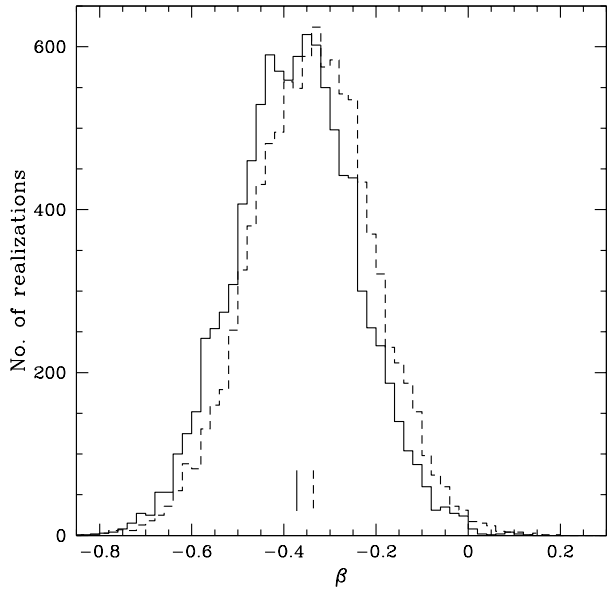} 
\caption{How do the SN distance errors affect the best-fit slope, 
$\beta$? The histograms show the distribution of derived $\beta$ values
in 10,000 realizations of the data. The solid (dashed) line histogram 
represents results using 50 (53) SNe. The short vertical lines at the
bottom of the plot indicate the medians of the distributions.}
\label{figextra}
\end{figure}

The best-fit line to the 50 SNe in Fig.~\ref{figone}
has a slope $\beta=-0.373$, and is shown as a thin slanting line. 
This fit does not include magnitude errors. To include the errors we do the 
following. We calculate the best-fit slope for 10,000 realizations of the data, 
where each data point's SN magnitude is replaced by a randomly picked magnitude
from a Gaussian distribution centred on the actual magnitude value, and having 
width equal to the quoted error. This procedure correctly incorporates the 
information contained in the errors, and produces a distribution of best-fit 
slopes, which is shown as a solid line histogram in Fig.~\ref{figextra}. 
This distribution shows that the median best-fit slope is $\beta=-0.372$, 
while $\beta=0$ (i.e. a case of no correlation) is ruled out at 99.80\% 
confidence level. Had we used 53 SNe (i.e. had we not exluded the 3 SNe with 
large distance errors), the median best-fit slope would gave been $\beta=-0.336$, 
while $\beta=0$ would have been ruled out at 99.37\% confidence level. The 
corresponding distribution of $\beta$ slopes is shown as the dashed line in 
Fig.~\ref{figextra}.

Next, we determine the likelihood of this relation arising by chance. To that
end, we estimate the significance of the relation in two ways. First, we assign 
random positions to SNe (keeping these control SNe on the same APM plate as
the original source), and redo all the analysis.
We repeat this 1000 times, and in 8 cases we find $\beta\le -0.373$, 
implying statistical significance of $99.2\%$. Second, 
we take the list of observed $M_{SN}$ values and randomly reassign them to 
observed SN sky positions. 10,000 randomized SNe samples are created in this 
fashion, and only $0.5\%$ of these have $\beta \le -0.373$.
Based on these two tests we conclude that the significance of the 
$M_{SN}$--$N_{>}/N$ relation is better than $99\%$.

These results are consistent with weak gravitational lensing,
which would amplify SNe found behind more nearby mass concentrations, as 
traced by APM galaxies. Alternatively, the results could be due to
the action of Galactic dust, which will obscure certain directions of
the sky making galaxies less numerous and SNe fainter. We consider 
Galactic dust further in Section~\ref{disc}; in Sections~\ref{model} and
~\ref{montecarlo} we proceed on the assumption that weak lensing is 
responsible to the $M_{SN}$--$N_{>}/N$ relation.

\section{Further tests}\label{robust}

The distribution of SNe points in Fig.~\ref{figone} depends on specific
choices that we made for certain parameters, in particular we chose circles
of radius $\theta=10^\prime$ and galaxy magnitude limit such that 
$z_{med}=z_l/x$, where $x=2$ (see Section~\ref{data}). How would the results
change if different choices were made? In other words, how robust is our
result, would it disappear had we picked a different set of parameters?

\begin{figure}
\includegraphics[width=84mm]{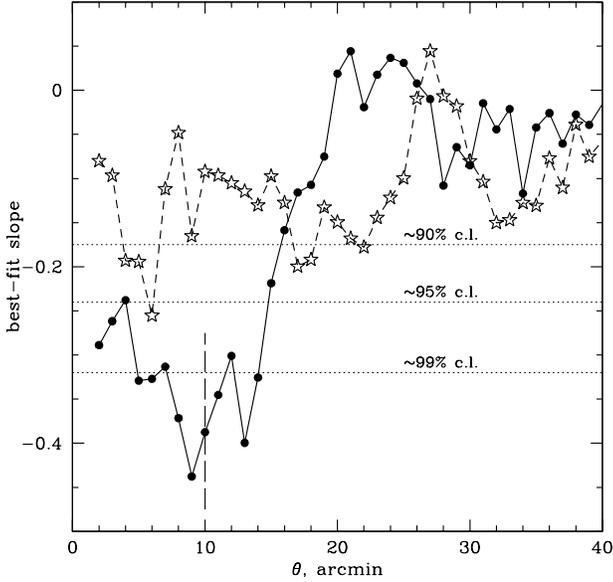}
\caption{Control tests: varying the aperture radius, $\theta$ of circles around SNe. 
The vertical axis shows $\beta$, the slope of the best-fit line to the relation
of the type shown in Fig~\ref{figone}. The filled circles
represent cases where galaxies were counted around SN; stars symbols represent
cases where Galactic stars were used instead (i.e. control experiment). 
The dotted lines denote, 
{\it approximately}, the levels of statistical confidence. The dashed vertical line
indicates radius $\theta$ used in the analysis of Section~\ref{anal}.}
\label{figthetatest}
\end{figure}

Figure~\ref{figthetatest} shows the effect of changing $\theta$. 
The vertical axis is $\beta$, the best-fit to the $M_{SN}$--$N_{>}/N$ relation
in each case.
Filled points represent cases where {\it galaxies} were counted and used to
determine the $N_{>}/N$ rank, while star symbols represent cases where
{\it Galactic stars} were used instead. As expected, $\delta n_{stars}$ do not
correlate with SNe magnitudes, and the values of the best-fit slopes 
are near zero. However,  because the APM star-galaxy classifier is
not perfect, some `stars' are actually galaxies, which accounts for some
signal being seen when using $\delta n_{stars}$. Horizontal dotted lines mark
the approximate location of $90\%, \,95\%$, and $99\%$ confidence levels.
These are only approximate because every point in the plot will have its
own significance level, but because the number of SNe contributing to each
point is the same in each case, and the total dispersion in SNe magnitudes
is the same, same $\beta$ values have about the same significance, 
regardless of $\theta$. We note that for very
small $\theta$ the galaxy numbers become very small, and Poisson noise drowns
out any $M_{SN}$--$N_{>}/N$ correlation that might exist, so the upturn
in the values of the best-fit slope at small $\theta$ is probably not real.
The dashed vertical line marks the $\theta$ value used in Section~\ref{anal}.
We conclude that significant $M_{SN}$--$N_{>}/N$ anti-correlations occur 
only with galaxies and not with Galactic stars (which serve as a control sample), 
and only for $5^\prime\simlt\theta\simlt 15^\prime$. 

\begin{figure}
\includegraphics[width=84mm]{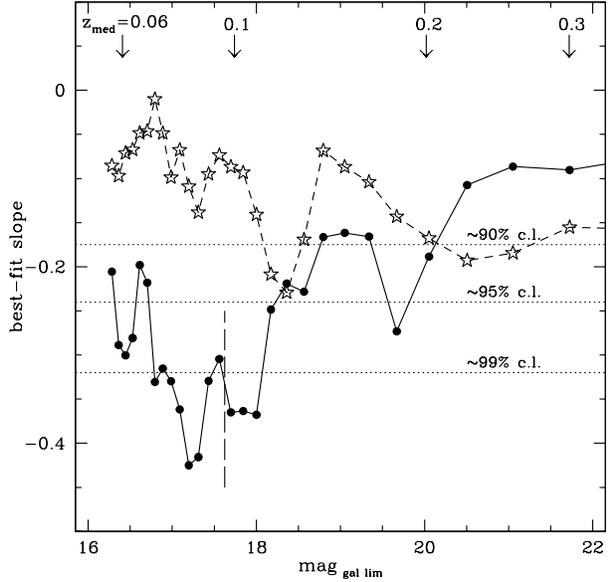}
\caption{Control tests: varying the apparent magnitude limit of galaxies
and the corresponding median redshift $z_{med}$. 
Similar to Fig.~\ref{figthetatest}; see Section~\ref{robust} for details.}
\label{figmagtest}
\end{figure}

Figure~\ref{figmagtest} shows the effect of changing the median redshift,
$z_{med}=z_l/x$ of the APM galaxies, or equivalently, mag$_{\;gal\;lim}$.
The hidden variable which is varied along the horizontal axis is $x$.
Since each SN has a different value of $z_l$ and hence 
mag$_{\;gal\;lim}$, depending on its $z_s$, there is no unique way of 
labeling the horizontal axis by using galaxy magnitudes, or redshifts.  
We label that axis by assuming $z_s=0.47$, 
the median of the SNe redshift distribution.  At the top of the plot we 
show how galaxy magnitude limit on the horizontal axis translates into the 
median galaxy redshift (for a specific case of $z_s=0.47$). 
We see that significant anti-correlation between $M_{SN}$ and $N_{>}/N$ of
foreground galaxies occurs for galaxies with $z_{med}\simlt 0.12$.
Using Galactic stars instead of galaxies produces no significant results. 
The value we used in Section~\ref{anal} is
shown with a vertical dashed line, and corresponds to $z_{med}\approx 0.1$.
As in Fig.~\ref{figthetatest}
the upturn in the best-fit slope values at bright mag$_{\;gal\;lim}$
is probably due to Poisson noise. 
The range of redshifts of APM galaxies that act as the best lenses for SNe are
$\sim 0.05-0.2$; more distant galaxies show no signal, as expected, if lensing
is the correct interpretation of the data, because more distant galaxies
are either too close to SNe in redshift or are actually at the same $z$.

A number of other tests have been carried out as well. For example, 
instead of using a $z_s$-dependent mag$_{\;gal\;lim}$, we tried fixed 
values of $17.5, ~18.0, ~18.5$, which gave, $\beta=-0.35, -0.31, -0.27$,
respectively, comparable to, but somewhat smaller than those seen in
Fig.~\ref{figmagtest}. This is not surprising: $z_s$-dependent galaxy
magnitude limits pick optimal lens redshifts for each source, thus 
maximizing the observed lensing signature.

We also reran the analysis with subsamples of the entire 50-source sample.
We split the SNe according to the teams: 
20 Supernova Cosmology Project SNIa gave a $M_{SN}$--$N_{>}/N$ anti-correlation 
significant at 95.8\%, while the corresponding significance level for the 
30 SNe from the High-$z$ Supernova Search Team is 96\%. 
Because the quality of UKST APM (Southern Hemisphere) plates is higher than
POSS APM (Northern Hemisphere) plates, and because there is some
overlap between plates along the equator, we used UKST plates whenever possible.
Redoing the analysis using only the 36 UKST SNe we get the best fit slope 
$\beta=-0.46$ at a significance level of 99.6\%, while the 14 POSS SNe had
$\beta=-0.11$ and the correlation was not significant, which is not surprising
given the size of this subsample.  
Splitting the whole sample into low and high redshift groups we get the 
following results: 25 SNe with $z_s<0.47$ have a slope $\beta=-0.37$ at 96.8\%, 
while 25 SNe with with $z_s\ge0.47$ have a slope $\beta=-0.23$ at 88.9\%.
These tests suggest
that the $M_{SN}$--$N_{>}/N$ relation has a physical origin (weak lensing or
Galactic dust), and is not an artifact arising form one subset of the data.

\section{Modeling the observations}\label{model}

In this Section we adopt the weak lensing interpretation of the 
$M_{SN}$--$N_{>}/N$ relation. Our simple lensing  model
does not use the individual values of $z_s$, and the optimal redshift 
distribution of the corresponding lenses. In lieu of these parameters we use
the lensing optical depth of the matter traced by the galaxies, $\kappa_0$.
For a fixed global geometry, $\kappa_0$ depends on $z_s$ and the redshift
distribution of the mass traced by the galaxies.
We assume that the APM galaxies faithfully trace the mass up to some redshift,
$z_t$, then
\begin{eqnarray}
\kappa_0= \Omega_m\; \rho_{crit}\; (c/H_0)~\hskip1.5in\nonumber\\
\hskip0.50in\times \int_0^{z_t} {{(1+z)^2} \over 
              {[\Omega_m(1+z)^3+\Omega_\Lambda]^{1/2}\; \Sigma_{crit}(z,z_s)}}
         \; dz,
\end{eqnarray}
In that case, projected fractional mass excess is $\delta n_{mass}=\kappa/\kappa_0$, 
where $\kappa$ is the convergence, with respect to a smooth Universe (filled beam) 
for the corresponding line of sight.
In the weak lensing regime source amplification $A\approx(1+2\kappa)$; $A=1$
for an unlensed source. The smallest value that $\kappa$ can attain
is minus the total optical depth, $-\kappa_0$, which corresponds to emptying out
all the mass along the line of sight from the observer to $z_t$.

In addition to the value for $\kappa_0$, our model has three ingredients, 
(1) mass distribution, i.e. a probability distribution (pdf) for $\delta n_{mass}$, 
which is related to $\kappa$ pdf,
(2) a biasing scheme, which relates $\delta n_{mass}$ to the projected 
fractional galaxy number density excess, $\delta n_{gal}$, and 
(3) the dispersion in the intrinsic magnitudes of SNIa's. 
Given this information we generate synthetic SNIa samples,  
with 50 sources each, then compute $N_{>}/N$ each SN, and the observed SN 
magnitude, $M_{SN}$. The slope of the best-fit line to the 
$M_{SN}$--$N_{>}/N$ relation, $\beta$ is then used to test how well a given 
model reproduces the observations. 

Because our model assumes the same $z_l$ distribution for all sources,
the value of $\beta$ to compare our models' predictions to should be the one
obtained by using a constant mag$_{\;gal\;lim}\approx 18$, i.e. $\beta\approx -0.3$
(see fourth paragraph in Section~\ref{robust}).

The specifics of the three model ingredients, and the associated model parameters 
are described later in this Section. The goal of the modeling is to determine 
what set of parameters can reproduce the observations, i.e. have $\beta\le -0.3$.
We are particularly interested in what effect biasing has on the results.
It is often suggested that the large amplitude of QSO-galaxy correlations
mentioned in the Introduction is, at 
least in part, due to the fact that biasing is not a simple linear, one-to-one
mapping from $\delta n_{mass}$ to $\delta n_{gal}$. 
QSO-galaxy correlation function ($\omega_{QG}$), galaxy autocorrelation 
($\omega_{GG}$) and the matter fluctuation power spectrum ($P_k$) are related by
\begin{equation}
\omega_{QG}\propto b^{-1} \omega_{GG} \propto b P_k,
\end{equation}
where $b$ is the biasing parameter. This 
means that $\omega_{QG}$ probes a {\it combination} of $b$ and $P_k$.
Thus, for a given $P_k$, $\omega_{QG}$ can be enhanced with
positive biasing, especially if it is non-linear on relevant spatial scales.

The important difference between weak lensing induced QSO-galaxy correlations 
and the effect on standard candles we have studied here is that in the latter case 
biasing plays a minor role. The slope of the $M_{SN}$--$N_{>}/N$ relation is most
sensitive to the mass distribution. Our modeling demonstrates 
this later; first, we described the specifics of the three model ingredients.

\subsection{Mass distribution}

\subsubsection{The standard mass distribution}\label{stdmass}

Based on the results from cosmological N-body simulations \citep{jai00}  
the shape of the probability distribution function (pdf) of $\kappa$ is very 
roughly Gaussian, but asymmetric,
with the most probable $\kappa$, which we call $\kappa_m$ being less than 0, 
i.e. most sources are deamplified. The tail of the $\kappa$ distribution 
extends to high positive values of $\kappa$. We use these published results 
of \citet{jai00} to construct an approximate shape for our $\kappa$ pdf:
\begin{equation}
p(\kappa)=\left\{
\begin{array}{l}
exp\;(-[\kappa-\kappa_m]^2/2\sigma_{\kappa_1}), \quad {\rm for}\quad \kappa > \kappa_m\\
exp\;(-[\kappa-\kappa_m]^2/2\sigma_{\kappa_2}), \quad {\rm for}\quad \kappa\le\kappa_m
\end{array}
\right.
\label{stdpdf}
\end{equation}
The pdf is a combination of two half-Gaussians, with two widths, 
$\sigma_{\kappa_1}$ and $\sigma_{\kappa_2}$, which describe the high $\kappa$ and
the low $\kappa$ sides of the pdf respectively. The location of the peak of the
pdf, $\kappa_m$ is adjusted such that the average $\kappa$ is 0, which implies
\begin{equation}
\kappa_m=\sqrt{2/\pi}~(\sigma_{\kappa_2}-\sigma_{\kappa_1}).
\end{equation}
$\kappa_m$ is always negative, since the skewness of the pdf dictates that
$\sigma_{\kappa_1}>\sigma_{\kappa_2}$. From the numerically computed pdfs
of \citet{jai00} we estimate that for a flat $\Omega_m=0.3$ model, where
the sources are at $z_s\approx 0.5-1$, and the smoothing scale is 
$\sim 5^\prime-10^\prime$, the following approximations apply:
$\sigma_{\kappa_1}/\sigma_{\kappa_2}\approx 2.6$,
and $\sigma_{\kappa_1}\approx \kappa_0/3.6$. 
Thus we now have a realistic $\kappa$ pdf,
appropriate for a standard cosmology with standard mass distribution. 
Figure~\ref{grid}(d) shows such a pdf as a short-dash line.

\subsubsection{Non-standard mass distribution: bifocal lens}\label{bifocalmass}

In addition to the standard $\kappa$ pdf described above,  
we also try an extreme form for $\kappa$ pdf obtained using the ``bifocal lens''
mass distribution proposed by \citet{kov91}:
\begin{equation}
p(A)={{1-A_2}\over{A_1-A_2}}\;\delta(A-A_1)+{{A_1-1}\over{A_1-A_2}}\;\delta(A-A_2)
\label{bifocals}
\end{equation}
where $\delta$ is the Kronicker's delta function, $A_1>1$, and $A_2<1$. 
For a fixed $\kappa_0$, this pdf can produce more pronounced lensing effects 
than the standard pdf, if $A_2$ is assigned the smallest allowable value, and
$A_1$ is made very large. This is because the standard pdf has a large probability
near $A=1$, while the bifocal pdf avoids such values altogether.

The distribution of mass corresponding to eq.~\ref{bifocals} is unrealistic.
One can make it somewhat realistic by allowing a range of $A_1$ and $A_2$ values. 
We do the following: for any one line of sight we randomly pick $A_1$ and $A_2$ from 
specified ranges, using flat priors, and eq.~\ref{bifocals} sets the probability
distribution of the two amplifications. We use two sets of ranges:
(1) $1<A_1\le (1+5\;\kappa_0)$ and $(1-\kappa_0)<A_2\le 1$ (we will call this model 
bifocal I model),
(2) $1<A_1\le (1+10\;\kappa_0)$ and $(1-2\;\kappa_0)<A_2\le 1$ (we will call this 
bifocal II).
The corresponding $\kappa$ pdfs, obtained by considering many lines of sight are
shown in Fig.~\ref{grid}(d) as a solid line for bifocal II and as a long-dash line 
for bifocal I. In bifocal II
the minimum value of $A_2$ is the minimum allowed value; the other limits on $A$'s
were picked arbitrarily. For comparison, the standard $\kappa$ pdf of 
Section~\ref{stdmass} is also shown, as the short-dash line. 
Optical depth of $\kappa_0=0.02$ was assumed for all three.
Compared to a standard pdf, bifocal pdfs imply that most of the lines of sight are 
rather empty, and there are a few lines of sight with very high values of $\kappa$.

\subsection{Biasing scheme}\label{biasscheme}

Projected fractional mass excess, $\delta n_{mass}$ is related to 
the observationally accessible quantity, $\delta n_{gal}$ through biasing.
We chose ``power law'' biasing, motivated by numerical simulations of 
\citet{co93} and \citet{dl99},
\begin{eqnarray}
(1+\delta n_{gal})=\hskip2.2in\nonumber\\
\left\{
\begin{array}{l}
(1+\delta n_{mass})^{\alpha_1}+b_s\;(1+\delta n_{mass}),
         \,\,{\rm for}\,\,\,\delta n_{mass}>0\\
(1+\delta n_{mass})^{\alpha_2}+b_s\;(1+\delta n_{mass}),
         \,\,{\rm for}\,\,\,\delta n_{mass}\le 0
\end{array}
\right.
\label{biasing}
\end{eqnarray}
where the power law part
allows the biasing to be non-linear, with different indexes depending on the sign 
of $\delta n_{mass}$, while $b_s$ is a stochastic biasing component which is chosen 
randomly for each SN; $b_s$ distribution has a Gaussian shape and width $\sigma_{b_s}$. 
Factor $(1+\delta n_{mass})$ multiplying $b_s$ ensures that the dispersion in
$\delta n_{gal}$ is reduced in underdense regions.
Qualitatively, the $\delta n_{gal}$ vs. $\delta n_{mass}$ relations produced by
eq.~\ref{biasing} look similar to those in Fig.1 of \citet{dl99}.

\subsection{Dispersion in the intrinsic SNIa magnitudes}

We assume that the dispersion in magnitude about the perfect standard candle case
has a Gaussian shape, with width $\sigma_{SN}$.

\section{Results of Monte Carlo simulations}\label{montecarlo}

Each model has five independent parameters: \\
$\{\kappa_0,\alpha_1,\alpha_2,\sigma_{b_s},\sigma_{SN}\}$.
We assume flat priors for the three biasing parameters, $\alpha_1$, $\alpha_2$, and
$\sigma_{b_s}$, as well as for $\sigma_{SN}$. 
Each specific set of parameters together with one of the three mass distribution
models generates a synthetic realization of the 50-source 
SNIa sample. From this entire ensemble of realizations, we only consider those 
that satisfy these observational constraints:
(1) the total rms dispersion in the synthetic SNIa magnitudes 
(which includes intrinsic and lensing induced contributions) must be within 0.05 mag
of the actual observed value of 0.3mag;
(2) the moments of the synthetic $\delta n_{gal}$ distribution 
(average, standard deviation, skewness and kurtosis) must be reasonably close to those 
of the observed distribution, which is characterized by (0.063, 0.52, 1.7, 4.9).
We chose ``reasonably close'' to mean that synthetic values must not be more than a
factor of 2 away from the actual values. 
Because of these constraints, the range from which the values of
$\alpha_1$, $\alpha_2$, $\sigma_{b_s}$, and $\sigma_{SN}$ parameters are picked 
is not relevant, as long as it not too restrictive. In other words, the observational
constraints eliminate cases with very large values of these parameters. 

The data suggests that demagnified SNe are not lost due to the flux limit. If they
were, we would have less SNe at small values of $N_{>}/N$, and more
at high values of $N_{>}/N$, whereas in the data (Fig.~\ref{figone}) the sources 
are roughly equally spread over the entire $N_{>}/N$ range. So our synthetic 
lensing models assume that we do not lose SNe because of deamplification.

\begin{figure}
\includegraphics[width=84mm]{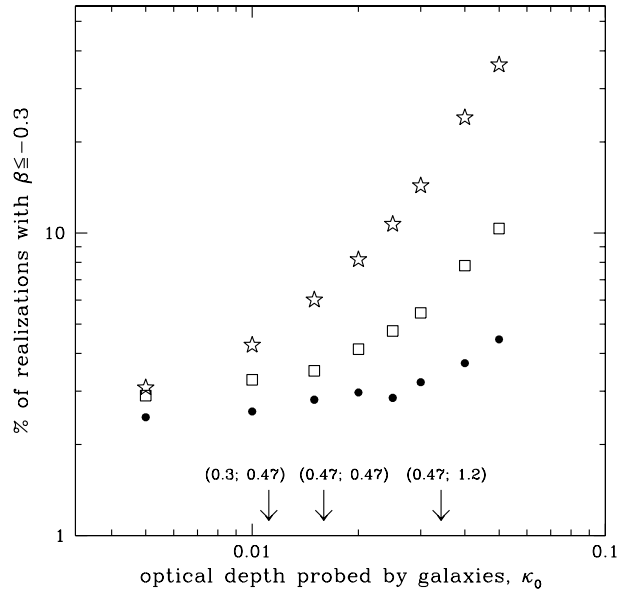}
\caption{Probability of obtaining the observed results 
(slope $\beta$ of the $M_{SN}$ vs. $N_>/N$ relation) 
given different mass distribution models, and a range of optical depths.
Solid dots assume standard $\kappa$ pdf of Section~\ref{stdmass}, 
empty squares and star symbols represent bifocal I and II models, respectively,
discussed in Section~\ref{bifocalmass}.
Each point was obtained using 10,000 synthetic realizations of SNIa 50-source
sample. See Section~\ref{montecarlo} for details.}
\label{figoptdepth}
\end{figure}

For each synthetic realization of the 50-source SNIa sample we derive the corresponding
$M_{SN}$--$N_{>}/N$ relation. The slope of the best-fit line, $\beta$ is recorded. 
For a given value of $\kappa_0$ we generate 10,000 realizations. 
Solid dots in Fig.~\ref{figoptdepth} show the percentage of realizations that 
have $\beta$ smaller than the observed value of -0.3, as a function of total 
optical depth, for the standard $\kappa$ pdf. 
Empty squares and star symbols represent the results for the
bifocal I and II. 

Figure~\ref{figoptdepth} considers a range of optical depths.
What is the appropriate value for $\kappa_0$?
If the source is at the median source redshift for our sample, $z_s=0.47$, then
APM galaxies sample the mass distribution up to a redshift of about 0.3. In this case
the lensing optical depth probed by APM galaxies is 0.011 (left-most arrow
in Fig.~\ref{figoptdepth}). A somewhat more optimistic estimate for $\kappa_0$ is
obtained if we assume that galaxies trace the mass fluctuations up to the
median source redshift, 0.47, which gives $\kappa_0=0.016$ (middle arrow). 
The limiting case is 
obtained by considering the most distant source, at $z_s=1.2$, and assuming that 
galaxies probe mass up to a redshift of 0.47, which gives $\kappa_0=0.034$ 
(right-most arrow).

\begin{figure}
\includegraphics[width=84mm]{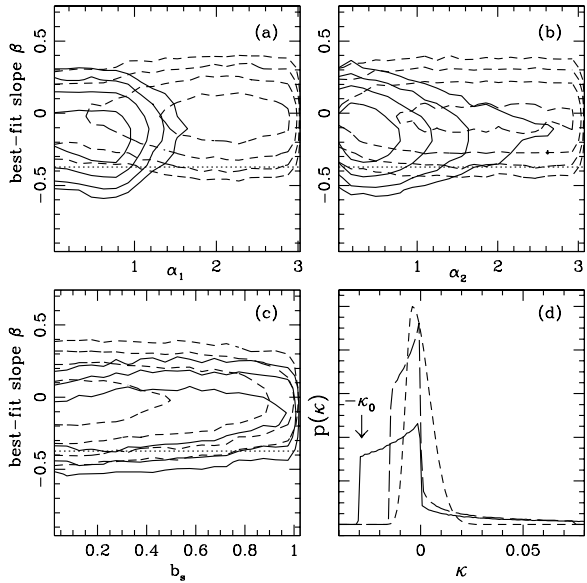}
\caption{~{\it Panels (a)--(c)} show the dependence of the best-fit slope of the 
$M_{SN}$--$N_>/N$  relation on the three biasing parameters 
(see Section~\ref{biasscheme}). Short-dash lines are for the standard $\kappa$ pdf 
model of Section~\ref{stdmass}, while the solid lines are for the bifocal II 
$\kappa$ pdf described in Section~\ref{bifocalmass}. Both models assume optical 
depth $\kappa_0=0.03$.  The contours are based on 50,000 realizations, and 
are spaced a factor of 3 apart. The shapes of the contours are partly determined 
by the observational constraints, described in Section~\ref{montecarlo} 
(second paragraph). It is evident that biasing parameters have little impact 
on the $\beta$. 
~{\it Panel (d):}~ Short-dash and solid lines represent $\kappa$ pdfs 
corresponding to the contours in panels (a)-(c) of the same line type. The 
minimum $\kappa$ in the bifocal II pdf is $-\kappa_0$ (indicated by an arrow).
The long-dash line represents bifocal I $\kappa$ pdf.
}
\label{grid}
\end{figure}

The important conclusion from this figure is that if the standard mass distribution 
(Section~\ref{stdmass}) is assumed, then the value of $\kappa_0$ does not 
really matter, and the probability of reproducing the observations is $\simlt 3\%$,
if $\kappa_0$ is within a reasonable range.
If, on the other hand, the bifocal pdf I or II are assumed, the probability of 
reproducing observations depends sensitively on the assumed optical depth. 
If $\kappa_0=0.034$ then these models yield 6-20\% probability.  
The bifocal pdfs produce more discernible lensing effects, for the same $\kappa_0$
because they have a wider range in $A$, as seen in Fig.~\ref{grid}(d).

The results presented in Fig.~\ref{figoptdepth} marginalize over values of all the 
model parameters (except $\kappa_0$), so the effects of the biasing parameters on 
$\beta$ are hidden. However, it turns out that these parameters do not correlate 
with $\beta$.
Figure~\ref{grid} shows the dependence of $\beta$ on $\alpha_1$, $\alpha_2$ 
and $b_s$ for
the bifocal II $\kappa$ pdf model (solid contour lines), and the standard $\kappa$
pdf model (dashed contour lines), both for $\kappa_0=0.03$. Fifty thousand 
realizations were created for each model, and the corresponding contour lines 
plotted, with adjacent contours separated by a factor of 3. 
It is apparent that for both models the dependence on the biasing parameters is weak. 

We conclude that {\it the slope of the $M_{SN}$--$N_>/N$ relation depends on the amount 
and distribution of matter, i.e. on $\kappa_0$ and the shape of the $\kappa$ pdf model, 
but does not depend on the specifics of the biasing scheme.}
Figure~\ref{figoptdepth} quantifies the dependence on amount and distribution of mass:
as expected, higher optical depth results in more pronounced lensing effects. 
Standard $\kappa$ pdf produces less lensing for the same $\kappa_0$ compared to
a much broader bifocal pdf. Figure~\ref{grid} demonstrates that biasing has little
effect on the slope of $M_{SN}$--$N_>/N$ relation. In other words, the type of
lensing signature considered here ($M_{SN}$--$N_>/N$ relation for standard candles) 
probes mass distribution, and is independent of biasing.

\section{Conclusions}\label{disc}

We detect the signature of weak lensing in the current sample of 50 high redshift SNIa
taken from two teams: High-$z$ Supernova Search Team and Supernova Cosmology Project.
After correcting SNe magnitudes for cosmological distances 
(assuming \citet{tonry03} values), we find that brighter SNe are
preferentially found behind regions overdense in foreground galaxies. This
$M_{SN}$--$N_>/N$ relation has a slope $\beta=-0.3$ to $-0.4$, when the angular radii
of foreground regions are 5-15 arcmin. The statistical significance is $>99\%$
(see Fig.~\ref{figthetatest} and ~\ref{figmagtest}).
The angular radii of 5-15 arcmin imply that the lensing structures are
$1-3\;{h_{0.7}}^{-1}$ Mpc across if they are located at a redshift of 0.1,
and so correspond to non-linearly evolved intermediate-scale structure.

Aside from the possibility that the observed $M_{SN}$--$N_>/N$ relation is a fluke, 
there is one other possible, non-lensing interpretation:
Galactic dust obscuration, which would make SNIa sources brighter and APM galaxies
more numerous in the directions devoid of Galactic dust. If this interpretation
is correct, then (1) Galactic extinction corrections applied to the SNe 
\citep{tonry03} are much too small, and (2) Galactic dust is able to 
change projected APM galaxy number density by $50\%$, since the observed rms 
in $\delta n_{gal}$ is $\approx 0.5$. 
We cannot comment on (1), but can estimate the effect of (2).
How much excess/deficit in galaxy density can Galactic dust create? Using
\cite{sfd98} data we calculate the average
$A_B$ Galactic extinction for the 50 SNe sample to be 0.18 mag. The slope
of the R-band APM galaxy number counts is $d\log N/dm=0.35$, therefore typical
``dust-induced''
$(1+\delta n_{gal})=10^{A_B\cdot d\log N/dm}\approx 1.16$. This is an upper bound
because extinction in the $R$ band will be smaller than in $A_B$, and because
extinction averaged over 5-15 arcmin radius patches will be smaller than estimates
for individual SNe, given that \cite{sfd98} extinction map resolution is 
6$^\prime$.1 FWHM.  If Galactic dust is indeed the cause of the $M_{SN}$--$N_>/N$
correlation, then $M_{SN}$ and $N_>/N$ should separately correlate with 
SNe extinction.  In reality, \cite{sfd98} extinction estimates do not show a 
correlation with either $M_{SN}$ or $N_>/N$. Overall, Galactic dust
interpretation of the observed $M_{SN}$--$N_>/N$ relation seems unlikely, 
but cannot be ruled out. 

Note that dust {\it intrinsic} to the $z_l\sim 0.1$ structures probed by the 
APM galaxies cannot be invoked to explain the observations, because such dust
would produce an effect opposite to the one detected here, i.e. $\beta$ would 
be positive. If dust is present in groups and clusters traced by the APM,
it will diminish the amplitude of the effect we detect. 
Presence of dust in groups was suggested by \citet{bfs88} who found that 
faint QSO candidates are anti-correlated with foreground groups. However,
such an anti-correlation can also be explained by weak lensing 
\citep{rwh94,cs99}. Current observations indicate that groups 
and clusters do not contain significant amounts of dust \citep{f93,nwm03}. 

If the lensing interpretation in correct, then the standard models of mass 
distribution have some difficulty in reproducing $\beta$; the observed 
value would be detected only in $\simlt 3\%$ of the cases. We investigate how
$\beta$ is affected by the amount and distribution of mass along the light of
sight to the sources, and galaxy biasing schemes. 

We find that larger optical depths and broader $\kappa$ pdf result in steeper
$\beta$ slopes (Fig.~\ref{figoptdepth}). Optical depth is a function of global 
geometry, and no realistic cosmological model can give $\kappa_0\simgt 0.05$, 
which is what would
be required to comfortably explain the results. Broader $\kappa$ pdfs mean that
mass fluctuations are more extreme than the standard cosmological models
allow, a scenario which is in apparant conflict with other means of determining
mass fluctuations, like cosmic velocity flows and weak shear lensing.

We also find that biasing has little effect on $\beta$ (Fig.~\ref{grid}).
This insensitivity to biasing is in contrast to weak lensing induced 
QSO-galaxy and GRB-galaxy (anti-)correlations, where biasing could, at least 
partly explain the higher than expected amplitude of the effect \citep{jai03,wf03}. 
We conclude that weak lensing of standard candles provides a cleaner probe of 
the mass distribution at $z\simlt 0.1$, on $\sim$few Mpc scales, then lensing 
induced angular correlations.

In fact, SNIa can provide the perfect means of measuring mass inhomogeneities
using gravitational lensing. A set of standard candles at known redshifts
can be analyzed using the complementary techniques of weak magnification and 
weak shear lensing. The advantage of magnification lensing over the more
commonly used shear lensing is that with the former one can chose 
the redshift range of the lenses, whereas the latter yields the cumulative
effect of lensing along the entire line of sight to the source. Therefore 
a large uniform set of intermediate redshift SNIa, such as the ones that would 
result from 
the SNAP mission\footnote{{\tt http://snap.lbl.gov}} and
the LSST mission\footnote{{\tt http://www.lsst.org/lsst$\_$home.html}}
would be invaluable for the studies of mass clustering in the nearby Universe. 
In addition to detecting intermediate and high-redshift SNe for the purposes
of estimating the global cosmological parameters and the equation of state
of the dark energy, SNAP will also measure weak shear lensing signature
due to large scale structure. Combining magnification data (of the type
considered in this paper) and shear information \citep{rho03}
for a large set of SNe will allow the study of mass distribution at $z\simlt 1.5$
with unprecedented accuracy and 3-dimensional spatial resolution.

The final issue we address is the impact of weak lensing on the
determination of global cosmological parameters, $\Omega_m$ and $\Omega_\Lambda$
using SNIa standard candles.
In principle, weak lensing can affect the derived values of 
$\Omega_m$ and $\Omega_\Lambda$, if (1) demagnified SNe are preferentially lost 
from the sample due to faint flux cutoff, and/or (2) the $\kappa$ pdf is asymmetric
and the SN sample size is small (see also \citet{wcxo97}).
If demagnified SNe were lost from
the sample then the distribution of SNe in $N_>/N$ would be skewed in the
direction of larger $N_>/N$ values. If $\kappa$ pdf is asymmetric,
(and it is, on scales considered here), then most SNe in a small sample 
will be slightly demagnified compared to average, and will have their
$N_>/N$ skewed in the direction of smaller values. So both
(1) and (2) would make the distribution of SNe in $N_>/N$ uneven, and would 
to some extent cancel each other. In the present sample of 50, the distribution 
of SNe in $N_>/N$ is indistinguishable from uniform, so the average $M_{SN}$ 
corresponds to the average $N_>/N$, and hence lensing effects by 
$\sim$Mpc-size structures probably did not bias the derived values of 
$\Omega_m$ and $\Omega_\Lambda$.

\section*{Acknowledgements}
The authors would like to thank Peter Garnavich and Jason Rhodes for
their careful reading of the manuscript and valuable comments.

\label{lastpage}

\end{document}